\documentstyle[12pt,aaspp4]{article}

\def\ges{\;_\sim^>\;}
\def\les{\;_\sim^<\;}

\slugcomment{Submitted to ApJ}

\lefthead{Guerra, Daly, \& Wan}
\righthead{Cosmological Parameters from Radio Galaxies}

\begin{document}

\title{Updated Estimates of Global Cosmological Parameters Determined 
Using Classical Double Radio Galaxies}

\author{Erick J. Guerra\altaffilmark{1}, Ruth A. Daly\altaffilmark{2,3},
and Lin Wan}
\affil{Department of Physics, Princeton
University, Princeton, NJ 08544}

\altaffiltext{1}{Present address: Department of Chemistry and Physics,
Rowan University, Glassboro, NJ 08028;\\ eddie@scherzo.rowan.edu}
\altaffiltext{2}{Present address: Department of Physics, Bucknell University,
Lewisburg, PA 17837; daly@bucknell.edu}
\altaffiltext{3}{National Young Investigator}

\begin{abstract}
A sample of 20 powerful extended radio galaxies with redshifts
between zero and two were used to determine constraints on 
global cosmological parameters.  Data for
six radio sources were obtained from the VLA archive,
analyzed, 
and combined with the sample of 14 radio galaxies used 
previously by Guerra \& Daly to determine cosmological
parameters.  The new results are consistent with our previous
results, and indicate that the current value of the 
mean mass density of the  universe
is significantly less than the critical value.  A universe with 
$\Omega_m$ in matter of unity is ruled out at 99.0\% 
confidence, and the best fitting values of $\Omega_m$ in matter
are $0.10^{+0.25}_{-0.10}$ and $-0.25^{+0.35}_{-0.25}$  assuming 
zero space curvature and zero cosmological
constant, respectively.  

The radio properties of each source are also used to determine
the density of the gas in the vicinity of the source, and
the beam power of the source.  
The six new radio sources have physical characteristics
similar to those found for the original 14 sources.  
The density of the gas around these radio sources
is typical of gas in present day clusters of galaxies.  
The beam powers are typically about $10^{45} \hbox{ erg s}^{-1}$.  

\end{abstract}


\section{Introduction}

The future and ultimate fate of the universe can be predicted 
given a knowledge of the recent expansion history of the
universe (assuming the universe is homogeneous and isotropic on 
scales greater than the current horizon size).  
This recent expansion history can be probed by 
studying the coordinate distance to sources at redshifts of
one or two; the coordinate distance is equivalent to the
luminosity distance or angular size distance multiplied by factors
of $(1+z)$.  The advantage of determining cosmological parameters
using the coordinate distance (luminosity distance or 
angular size distance) is that this distance depends on
global, or average, cosmological parameters.  It is independent
of the way the matter is distributed spatially, of the 
power spectrum of density fluctuations, of whether the matter
is biased relative to the light, and of the form or nature
of the dark matter (assuming that the universe is homogeneous
and isotropic on large scales).  

It was shown in 1994 that powerful double-lobed radio galaxies provide
a modified standard yardstick that can be used to determine
global cosmological parameters (Daly 1994, 1995), much like supernovae
can be used as modified standard candles.  The method was 
applied and discussed in detail by Guerra \& Daly (1996, 1998), Guerra
(1997), and Daly, Guerra, \& Wan (1998) who found that the data 
strongly favor a low density universe; a universe with $\Omega_m
=1$ was ruled out at 97.5 \% confidence.    

It was shown by Daly (1994, 1995) that the radio properties of
these
sources could be used not only to study global cosmological
parameters, but also to determine the ambient gas density, 
beam power, Mach number of lobe advance, and ambient gas
temperature of the sources and their environments.  The 
characteristics of the sources and their environments are 
presented and discussed in a series of papers (Wellman \& Daly
1996a,b; Wan, Daly, \& Wellman 1996; Daly 1996;  
Wellman, Daly,
\& Wan 1997a,b; Wan \& Daly 1998a,b; Wan, Daly, \& Guerra 1998).  

Radio maps of six powerful double-lobed radio galaxies were extracted from the 
Very Large Array (VLA) archives 
at the National Radio Astronomy Observatory (NRAO), and analyzed in detail.  
New results on global cosmological parameters, ambient gas
densities, and beam powers are presented here.  

The expanded sample is described in \S 2.  The new results are
presented in \S 3.   The implications of the results are
discussed in \S 4.  

\section{Expanded Sample}\label{sec:samp}

Each powerful extended radio galaxy (also known as a ``classical double'') 
has a characteristic size, $D_*$, that predicts the lobe-lobe
separation at the end of its lifetime 
(Daly 1994, 1995; Guerra \& Daly 1998).
The parameters needed to compute the characteristic size
are the lobe
propagation velocity, $v_L$, the lobe width, $a_L$, and the
lobe magnetic field strength, $B_L$.
These three parameters can be determined using 
radio maps with arc-second resolution at multiple frequencies,
such as those produced with the 
VLA or MERLIN (Multi-Element Radio Linked Interferometer Network).  
Multiple-frequency data
are needed to use the theory of spectral aging to estimate
the lobe propagation velocity (e.g., Myers \& Spangler 1985).  In addition, 
these maps
must have the necessary angular resolution and dynamic
range to image sufficient portions of the radio bridges.

Two published
data sets, Leahy, Muxlow, \& Stephens
(1989) and Liu, Pooley, \& Riley (1992),
have radio maps of powerful extended radio galaxies
at multiple frequencies which are sufficient
to compute all three parameters used in determining $D_*$.
These data were used
to compute $D_*$
for 14 radio galaxies
(Guerra \& Daly 1996, 1998; Guerra 1997; Daly, Guerra, \& Wan 1998).
Current efforts to expand the data set are underway, and include
searches through the VLA archive.  The VLA archive search
has already yielded the desired data for six radio galaxies, and new
results including these six sources are presented here.

Data were selected from observations
of powerful extended radio galaxies from the 3CR sample 
(Bennett 1962) on the basis of the observation frequency and
array configuration used.  An observation in the VLA archive was
a candidate if the lobe-lobe angular size of the source was 10 to 40
times the implied beam size, and four hours separated the first and 
last scans.  These observations should
resolve the source sufficiently and have enough $uv$-coverage
to image the bridges.  From candidate observations at
both L and C band, we have 
successfully imaged data from six sources.  The VLA 
archive data sets used here
are listed in Table \ref{tab:arch}.  


Radio imaging was performed using the NRAO
${\cal AIPS}$ software package.
The $uv$ data needed minimal editing, and initial calibration
was performed in the standard manner using ${\cal AIPS}$. 
As diagnostics, initial images were made using the ${\cal AIPS}$
task {\tt IMAGR} both without and with the {\tt CLEAN}
deconvolution algorithm.  The final radio maps 
were produced with the {\tt SCMAP} task, which performs
self-calibration along with the 
{\tt IMAGR} and {\tt CLEAN} tasks.    

Parameters 
used in ${\cal AIPS}$ for imaging, deconvolution, and calibration are
chosen to produce radio maps for a given source that can 
easily be used for spectral aging analysis.
This is particularly important where observations
at different frequencies are produced by different observers.
Although some of these radio maps exist in published form,
different choices of parameters (such as the restoring beam)
are often used
to produce these radio maps.  Thus, reducing
the raw $uv$ data insures that the data analysis can
be performed consistently between radio maps.

The computation of $a_L$, $B_L$, and $v_L$ were performed
here in the same manner as Wellman, Daly, \& Wan (1997a,b).
The deconvolved lobe width, $a_L$, is measured $10 h^{-1}$ kpc
from the hot spot toward the host galaxy.  The lobe magnetic
field strength,
$B_L$, is computed from the deconvolved surface brightness and
bridge width measured $10 h^{-1}$ kpc
from the hot spot toward the host galaxy.  
The lobe propagation velocity, $v_L$,
is computed on the basis of spectral aging along the
imaged portions of the bridge, and the magnetic field used in spectral
aging is computed using values measured $10 h^{-1}$ kpc
and $25 h^{-1}$ kpc from the hotspot.
In this paper, all parameters are computed with 
$b=0.25$ which means
magnetic fields are computed to be 0.25 times 
the minimum energy values, and without an $\alpha-z$ correction
which refers to a correction related to 
the observed correlation between spectral index and redshift
(see Wellman 1997;
Wellman, Daly, Wan 1997a,b; Guerra 1998; Guerra \& Daly 1998 for details).
It was found by Guerra \& Daly (1998) that constraints on
cosmological parameters did not depend on these choices, and very
similar results are obtained independent of the value of $b$ and of
whether an $\alpha - z$ correction is applied.  

These three parameters are computed for each bridge, with the exception
of one bridge in 3C 324 which was not imaged along its length sufficiently.
The characteristic core-lobe size, $r_*$, was computed for each bridge using
using the equation (see Guerra \&
Daly 1998):
\begin{equation}
r_* \propto \left({1 \over B_L~a_L} \right)^{2 \beta/3}~
v_L^{1 - \beta/3}~~.
\end{equation}
The characteristic 
(lobe-lobe) size, $D_*$, is taken
to be the sum of both $r_*$ (or in the case of 3C 324,
$D_*= 2 r_*$), and is normalized so that $D_*$ of Cygnus A (3C 405), a very
low redshift source in our sample, is equal to the average size of the 
full population of powerful classical double radio galaxies at
very low redshift.  Constraints on cosmological parameters
determined using this method are independent of the
normalization of $D_*$.  
As discussed in \S \ref{ssec:cosmo}, the best fit value of the one
model parameter $\beta$ is determined simultaneously with the 
best fit values for the two cosmological parameters that enter,
$\Omega_m$ and $\Omega_{\Lambda}$, the normalized values of the
current values of the mean mass density and the cosmological
constant (see Guerra \& Daly 1998;
Daly, Guerra, \& Wan 1998).  


Table \ref{tab:dstab} presents the six new $D_*$ values
in the last column, assuming the best fit value of $\beta=1.75$
(see \S \ref{ssec:cosmo}).  
Source name and redshift are listed in the first two columns.
The third column lists the redshift bin corresponding to 
the assignments in Guerra \& Daly (1998) and Table \ref{tab:phys}
below. The lobe-lobe angular size of the source is listed in the fourth column,
and the fifth and sixth columns list the core-lobe characteristic sizes,
$r_*$.  

\section{Updated Results}\label{sec:res}

\subsection{Constraints on Cosmological Parameters}\label{ssec:cosmo}

The subsample of sources
with estimates of the characteristic size, $D_*$, has been increased
from 14 to 20, as discussed above in \S \ref{sec:samp}.  
Only sources with physical sizes, defined as the 
projected separation between the radio hot spots, greater than
$20\,h^{-1}$ kpc can be used to determine a characteristic size.  
This is because smaller sources are typically not sufficiently resolved
that the radio data are useful, and 
the radio lobes of smaller sources are
interacting with the interstellar medium of the host galaxy
rather than the intergalactic/intracluster medium.  
It was decided that this same criterion should be applied to 
the larger comparison sample of powerful 3CR radio galaxies. 
Thus, 
the sample of radio galaxies used to determine
the redshift evolution of the physical size
has been reduced from 82
to 70; twelve radio galaxies were cut because
the physical separation between their lobes was less than
$20\,h^{-1}$ kpc.  
This has a rather small impact on the actual means
and standard deviations of the parent population, as can be seen
by comparing Table \ref{tab:phys} of this paper with Table 1 of Guerra \&
Daly (1998).  
The average lobe-lobe separations as a function of redshift
are listed in Table \ref{tab:phys} for three example choices
of cosmological parameters (matter-dominated, curvature-dominated, and
spatially flat with non-zero cosmological constant). 

\begin{figure}
\plotone{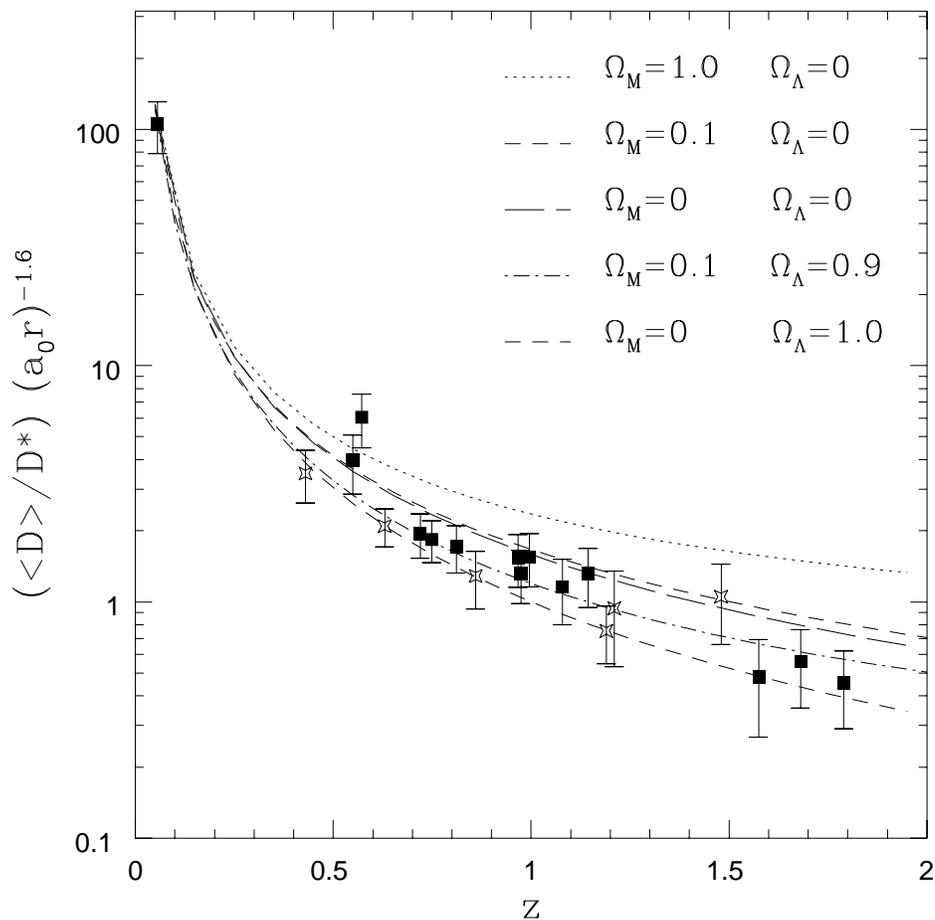}
\caption{The quantity $(\langle D \rangle / D_*) (a_o r)^{-1.6}$,
computed assuming $\beta=1.75$, $b=0.25$ and not including $\alpha-z$
correction. This measured quantity is independent $\Omega_m$ and
$\Omega_{\Lambda}$.
Open Stars denote the six new
data points from the VLA archive.
For different choices of $\Omega_m$ and $\Omega_{\Lambda}$, 
the predicted redshift evolution of 
$(a_o r)^{-1.6}$
is plotted for comparison. 
\label{fig:lin}}
\end{figure}

To solve simultaneously for the model parameter $\beta$ and the 
cosmological parameters $\Omega_m$ and $\Omega_{\Lambda}$, the
ratio of $D_*$
for each source to $\langle D \rangle$, the average lobe-lobe
size of the parent population in the
corresponding redshift bin,
is fit to a constant, independent of redshift. 
The value of the constant is a free parameter,
so the normalization of Cygnus A does not affect the
fits in any way.  
Figure \ref{fig:lin} illustrates the cosmological
dependence of $\langle D \rangle / D_*$ on the coordinate distance
$(a_o r)$.  For $\beta = 1.75$, $\langle D \rangle / D_*$ is
proportional to $(a_or)^{1.6}$; thus
$(\langle D \rangle / D_*)(a_or)^{-1.6}$ is independent of
cosmological parameters.  The data can be compared to
several different sets of cosmological parameters on a single
figure by plotting $(\langle D \rangle / D_*)(a_or)^{-1.6}$ for
each data point and comparing this with $(a_or)^{-1.6}$ curves
obtained for different sets of cosmological parameters, as is
shown in Figure \ref{fig:lin}.  

The hypothesis is that, for the 
correct choice of cosmological parameters, $\langle D \rangle /
D_* = 1$, so that the values
of $(\langle D \rangle / D_*)(a_or)^{-1.6}$ for 
all 20 radio galaxies should follow a curve that, at each z, is parallel 
to the curve $(a_or)^{-1.6}$ obtained for that particular choice of
cosmological parameters.  Figure \ref{fig:lin}, shows $(\langle D \rangle /
D_*)(a_or)^{-1.6}$ for the the six new
points and 14 original points as a function of z
(the six new points are denoted with stars).  
Also drawn on this figure are curves of $(a_or)^{-1.6}$ 
obtained for specified values of cosmological parameters.  In
this figure, all of the curves pass through Cygnus A, though
this is not required when we actually solve for best fitting
cosmological parameters.  It is clear that curves obtained for a 
low density universe describe the data points quite well, and the 
curve describing a universe with $\Omega_m = 1$ does not follow 
the data points.  
Note that actual fits are performed with a free normalization
that does not require the predicted curves to pass through
Cygnus A, the lowest redshift point.  Figure \ref{fig:lin} is
for illustrative purposes and is not part of any fits.


For all 20 sources, the chi-squared for fitting
the ratio to a constant is computed for relevant values of 
$\beta$, $\Omega_m$, and $\Omega_{\Lambda}$.  It is found 
that the best fit value of $\beta$ is $\beta = 1.75 \pm 0.25$. 
This result is insensitive to the choice of $\Omega_m$
and $\Omega_{\Lambda}$, and there appears to be no significant
covariance between $\beta$ and cosmological parameters
(see Figures \ref{fig:ombet}a,b). 


The confidence contours in the $\Omega_m$ - $\Omega_{\Lambda}$
plane are shown in 
Figures \ref{fig:2d} and \ref{fig:1d}.
The probability
associated with a given range of
$\Omega_m$ and $\Omega_{\Lambda}$ independent of $\beta$ is
shown in Figure \ref{fig:2d} (referred to as two-dimensional
confidence intervals). 
In Figure \ref{fig:1d}, the projection of a confidence interval
onto either axis ($\Omega_m$ or $\Omega_{\Lambda}$) 
indicates the probability associated with a given
range of that one parameter, independent of
all other parameter choices (referred to as one-dimensional confidence
intervals).
Both figures illustrate how this method and the data
are most consistent with a low density universe; $\Omega_m \les 0.15$
with 68\% confidence, $\Omega_m \les 0.5$ with 90\% confidence,
and $\Omega_m \les 1.0$ with 99\% confidence.
The constraints on $\Omega_{\Lambda}$ are not as strong, and values
of $\Omega_{\Lambda}$ from zero to unity are consistent
with the data.



The best fit value of $\beta = 1.75 \pm 0.25$ is consistent with 
the previous estimates of $\beta \simeq 1.5 \pm 0.5$ (Daly 1994), and
$\beta \simeq 2.1 \pm 0.6$ (Guerra \& Daly 1996, 1998), but with
significantly reduced uncertainties. Similarly, the constraints on 
cosmological parameters are consistent with previous estimates, 
(Daly 1994; Guerra \& Daly 1996, 1998; Guerra 1997; Daly, Guerra, \& Wan 1998) 
but with smaller error bars.  
It is apparent in Figures \ref{fig:2d} and \ref{fig:1d} that 
these data and method
strongly favor a low density universe; a universe where 
$\Omega_m = 1$ is ruled out with 99.0 \% confidence
independent of $\Omega_{\Lambda}$ and $\beta$ . 
This will be discussed in more
detail in \S \ref{sec:disc}.

\subsection{Ambient Gas Densities and Beam Powers}\label{ssec:other}

Daly (1994, 1995) showed that the radio properties of a powerful
extended radio source could be used to estimate the beam power,
$L_j$, and density of the gas in the vicinity of the source,
$n_a$.  The method 
is described in detail by Wellman (1997), Wellman, Daly, \& Wan 
(1997a), Wan (1998),
and Wan, Daly, \& Guerra (1998).  Values for the ambient
gas density and beam power for the 14 sources in the original
sample are described in these papers.  
The basic equations are:
\begin{equation}
L_j \propto a_L^2~B_{10}^2~v_L~~,
\end{equation} 
and
\begin{equation}
n_a \propto {B_{10}^2 \over v_L^2}~~; 
\end{equation}
the normalizations are
given in the references listed above. The values obtained for the six new
radio galaxies in our sample are listed in Table \ref{tab:natab}.  Also
listed
in Table \ref{tab:natab} are all the input parameters 
used to compute $L_j$, $n_a$, and $D_*$.


\section{Discussion}\label{sec:disc} 

A parent population of 70 powerful extended classical double
radio galaxies with redshifts between zero and two was used to 
define the evolution of the mean or characteristic size of these
sources as a function of redshift.  An independent estimate of
the mean or characteristic size of a given source was possible
for a subset of 20 of these radio galaxies for which extensive
multiple frequency radio data was available.  Requiring that the two 
measures of the mean source size have the same redshift behavior
allows a simultaneous determination of three parameters: the one
model parameter $\beta$, and the two cosmological parameters
$\Omega_m$ and $\Omega_{\Lambda}$ (assuming that the only
significant cosmological parameters today are the mean mass
density, a cosmological constant, and space curvature).  

\begin{figure}
\plotone{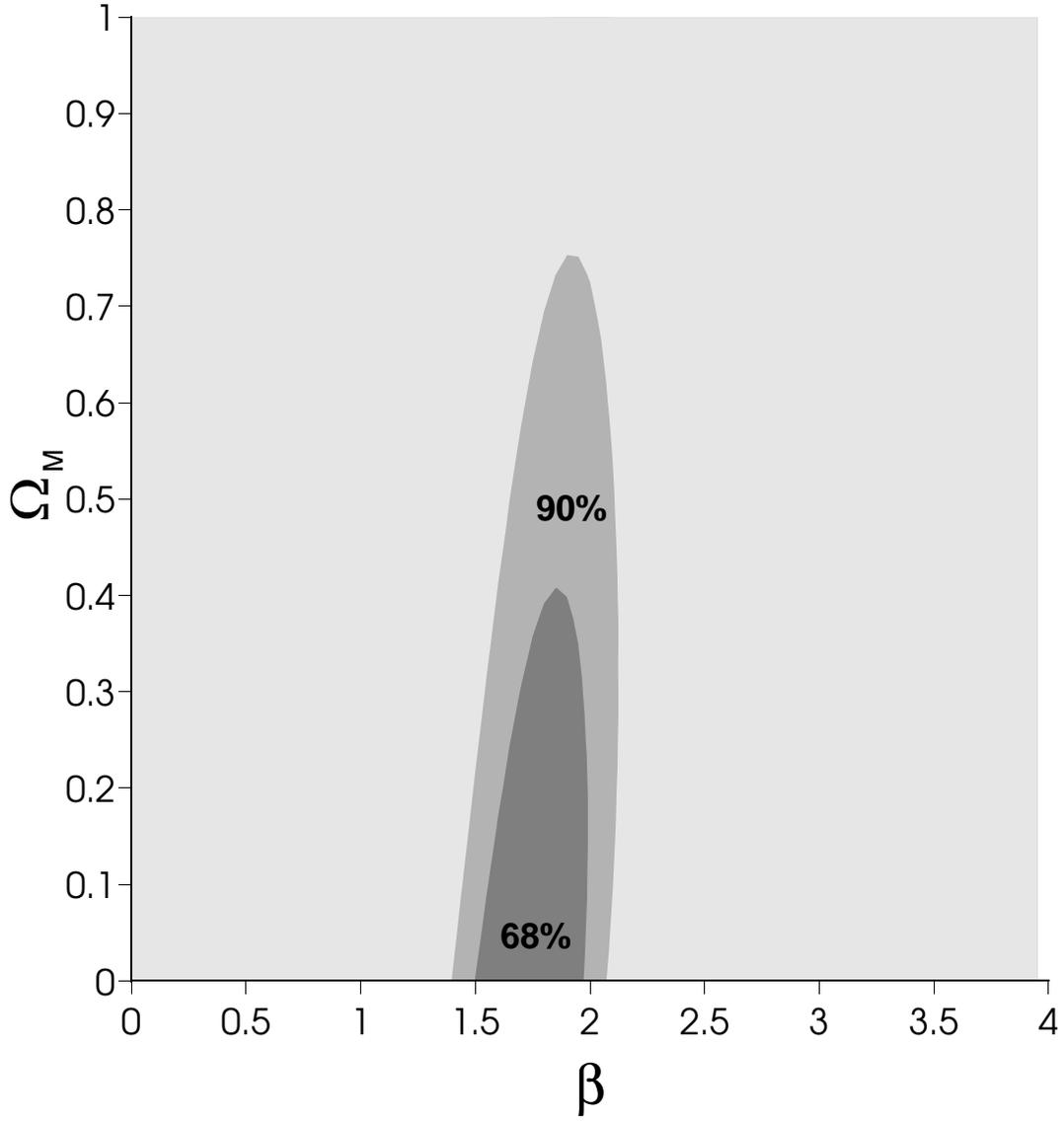}\\
\caption{The 68\% and 90\% confidence intervals for 
$\beta$ and $\Omega_m$, where (a) $\Omega_{\Lambda}$=0 (no cosmological
constant)...
\label{fig:ombet}}
\end{figure}

The method was applied to this data set, and interesting new constraints are
presented.  It is found that the model parameter is very tightly
constrained to be $\beta = 1.75 \pm 0.25$ 
(see Figure \ref{fig:ombet}), consistent with 
previous estimates, and that this model parameter is independent
of cosmological parameters.  For a value of $\beta = 1.75$, the 
characteristic source size is $D_* \propto
(B_La_L)^{-7/6}~{v_L}^{5/12}$, which indicates that it is necessary
to have multiple-frequency radio data in order to estimate
$D_*$, owing to its $v_L$ dependence.  

The data strongly favor a low density universe; a universe with 
$\Omega_m = 1$ is ruled out with 99\% confidence, independent
of the value of $\Omega_{\Lambda}$ or $\beta$.  Either space
curvature or a cosmological constant, or both, are allowed.  
The main conclusion is that $\Omega_m$ is low, but, at this
point, the method and data do not allow a discrimination between 
whether space curvature or a cosmological constant is important 
at the present epoch.  

It is interesting to note that the lowest
reduced chi-squared obtained is 0.96 for $\Omega_m \simeq -0.25$
and $\Omega_{\Lambda} \simeq 0$.
This value is slightly
greater and closer to unity than the minimum reduced chi-squared
obtained by Guerra \& Daly (1998), which indicates that
this sample of 20 sources has a reasonable distribution
around any model predictions.  This convergence
to unity with increasing sample size suggests
that this method and its statistics are reliable.

The best fit for cosmological parameters in the
physically relevant half-plane of $\Omega_m \ges 0$
is $\Omega_m = 0$ and $\Omega_{\Lambda} = 0.45$ with a
reduced chi-squared of 0.98 (also close to unity).
However,
Figures \ref{fig:2d} and \ref{fig:1d} clearly show that
our results are still consistent $\Omega_{\Lambda} = 0$.

The radio data also allow a determination of the density of the 
ambient gas in the vicinity of each radio source, and the beam
power of each source; the values of these quantities are
presented.  The sources lie in high-density gaseous environments
like those found in low-redshift clusters of galaxies.  Typical
beam powers for the sources are $\sim 10^{45} \hbox{ erg
s}^{-1}$.  

\acknowledgments

Special thanks go to Rick Perley for his aid in extracting 
data from the VLA archive.
The authors would also like to thank Katherine Blundell, 
Chris Carilli, Miller Goss, Paddy Leahy, Wil van Breugel,
and Dave Wilkinson for helpful discussions.  It is a pleasure to
acknowledge Ingrid Stairs, Rick Balsano, and other members of the pulsar group
who graciously allowed us to use their computer facilities.  
This work was supported in part by the U.S. National Science
Foundation,
the Independent College Fund of New Jersey, and a grant from 
W. M. Wheeler III.

\begin{deluxetable}{lcccccc}
\small
\tablecaption{\label{tab:arch}VLA Archive Data Sets}
\tablehead{
\colhead{} & \colhead{} & \colhead{} & \colhead{Listed} &
\colhead{Frequency} & \colhead{Array} & \colhead{} \\
\colhead{Source} & \colhead{Band}  & \colhead{Program ID} & 
\colhead{Observer} 
& \colhead{(GHz)} & \colhead{Config.} & \colhead{Date}  
}
\startdata

3C 244.1 & L & POOL & Pooley, G. & 1.411 & B & 09/17/82 \nl
 & C & AF213 & Fernini, I. & 4.872 & B & 12/23/91 \nl
3C 337  & L & AR123 & Rudnick, L. & 1.452 & A & 02/21/85 \nl
 & C & AR123 & Rudnick, L. & 4.885 & B & 06/01/85 \nl
3C 325  & L & AV153 & van Breugel, W. & 1.465 & A & 12/05/88 \nl
 & C & AF213 & Fernini, I. & 4.885 & B & 12/23/91 \nl
3C 194  & L & AV164 & van Breugel, W. & 1.465 & A & 05/11/90 \nl
 & C & AV164 & van Breugel, W. & 4.885 & A & 05/11/90 \nl
3C 324  & L & AR123 & Rudnick, L. & 1.452 & A & 02/21/85 \nl
 & C & AR123 & Rudnick, L. & 4.885 & B & 06/01/85 \nl
3C 437  & L & AV164 & van Breugel, W. & 1.465 & A & 05/11/90 \nl
 & C & AV164 & van Breugel, W. & 4.885 & B & 04/21/89 \nl

\enddata
\end{deluxetable}

\begin{deluxetable}{llccccc}
\small
\tablecaption{\label{tab:dstab}Radio Galaxies with $D_{\star}$ Presented Here}
\tablehead{
\colhead{}  & \colhead{}  & \colhead{}  &
\colhead{$\theta$}      &
\multicolumn{2} {c} {$r_{\star}$ \tablenotemark{a}}     &
\colhead{$D_{\star}$\tablenotemark{a}}  \nl
\cline{5-6}
\colhead{Source}        &
\colhead{$z$}   &
\colhead{Bin}   &
\colhead{(arcsec)}      &
\colhead{($h^{-1}$ kpc)}        &
\colhead{($h^{-1}$ kpc)}        &
\colhead{($h^{-1}$ kpc)}        }
\startdata

 3C 244.1   & 0.43 & 2        & 50.8  & $151\pm16$    & $135\pm16$    & $286\pm22$           \nl
 3C 337     & 0.63 & 3        & 43.5  & $150\pm24$    & $65\pm7$      & $214\pm25$           \nl
 3C 325     & 0.86 & 3        & 15.8  & $164\pm52$    & $66\pm14$     & $230\pm54$           \nl
 3C 194     & 1.19 & 4        & 14.2  & $105\pm22$    & $92\pm15$     & $197\pm27$           \nl
 3C 324     & 1.21 & 5        & 10.2  & $75\pm18$     & \nodata       & $149\pm35$           \nl
 3C 437     & 1.48 & 5        & 36.7  & $55\pm6$      & $48\pm5$      & $103\pm8$            \nl

\enddata
\tablenotetext{a}{Computed assuming $\beta=1.75$, $\Omega_0=0.1$,
$\Omega_{\Lambda}=0$, $b=0.25$ and not including $\alpha-z$
correction.}
\tablenotetext{b}{$r_{\star}$ for only one bridge.}
\end{deluxetable}

\begin{deluxetable}{ccccccc}
\small
\tablecaption{The Average Lobe-Lobe Sizes for 
Powerful 3CR Radio Galaxies. \label{tab:phys}}
\tablehead{
\colhead{} & \colhead{} & \colhead{} & \colhead{} & 
\multicolumn{3}{c}{$\langle D \rangle$ ($h^{-1}$ kpc)} \\  
\cline{5-7} \\
\colhead{Bin} & \colhead{$z$ Range} & \colhead{Sources} & \colhead{} &
\colhead{$\Omega_o=1.0$, $\Omega_{\Lambda}=0.0$} &
\colhead{$\Omega_o=0.1$, $\Omega_{\Lambda}=0.0$} &
\colhead{$\Omega_o=0.1$, $\Omega_{\Lambda}=0.9$} 
}
\startdata

1 & 0.0-0.3 & 3  & & 66$\pm$14  & 68$\pm$14  & 72$\pm$13  \nl
2 & 0.3-0.6 & 13 & & 202$\pm$45 & 224$\pm$50 & 259$\pm$57 \nl
3 & 0.6-0.9 & 23 & & 148$\pm$17 & 174$\pm$20 & 209$\pm$24 \nl
4 & 0.9-1.2 & 16 & & 107$\pm$24 & 133$\pm$29 & 165$\pm$36 \nl
5 & 1.2-1.6 & 9  & & 91$\pm$32  & 122$\pm$43 & 152$\pm$53 \nl
6 & 1.6-2.0 & 6  & & 66$\pm$19  & 92$\pm$26  & 114$\pm$33 

\enddata
\end{deluxetable}

\begin{deluxetable}{llcccccc}
\small
\tablecaption{\label{tab:natab}Summary of New Source Properties}
\tablehead{
\colhead{Source}        &
\colhead{$z$}   &
\colhead{$a_L$\tablenotemark{a}}        &
\colhead{$B_{10}$\tablenotemark{b}}     &
\colhead{$B_{25}$\tablenotemark{c}}     &
\colhead{$v_L$\tablenotemark{d}}        &
\colhead{$n_a$\tablenotemark{e}}        &
\colhead{$\log L_j$\tablenotemark{f}}
}
\startdata

 3C 244.1  & 0.43  & $5.5\pm0.3$   & $3.5\pm0.2$  & $2.5\pm0.2$   & $1.1\pm0.2$
 & $1.0\pm0.4$   & $44.35\pm0.10$         \nl
           &       & $4.7\pm0.4$   & $5.3\pm0.3$  & $3.2\pm0.2$   & $1.7\pm0.3$
 & $1.0\pm0.4$   & $44.75\pm0.11$         \nl
 3C 337    & 0.63  & $6.2\pm0.5$   & $3.5\pm0.2$  & $2.3\pm0.1$   & $1.5\pm0.5$
 & $0.6\pm0.4$   & $44.57\pm0.15$         \nl
           &       & $10.5\pm0.4$  & $3.8\pm0.2$  & $2.8\pm0.2$   & $1.1\pm0.2$
 & $1.2\pm0.4$   & $44.99\pm0.10$         \nl
 3C 325    & 0.86  & $3.3\pm1.1$   & $6.8\pm0.9$  & $4.0\pm0.3$   & $2.0\pm0.5$
 & $1.2\pm0.6$   & $44.73\pm0.23$         \nl
           &       & $4.4\pm0.8$   & $12.7\pm1.0$ & $4.6\pm0.3$   & $3.0\pm0.8$
 & $1.9\pm1.0$   & $45.70\pm0.17$         \nl
 3C 194    & 1.19  & $4.5\pm1.0$   & $7.1\pm0.6$  & $5.0\pm0.3$   & $1.8\pm0.3$
 & $1.6\pm0.6$   & $45.00\pm0.16$         \nl
           &       & $5.6\pm0.8$   & $6.7\pm0.5$  & $5.6\pm0.4$   & $2.0\pm0.3$
 & $1.2\pm0.4$   & $45.16\pm0.13$         \nl
 3C 324    & 1.21  & $4.5\pm0.9$   & $10.3\pm0.9$ & \nodata       & $2.1\pm0.7$
 & $2.4\pm1.5$   & $45.38\pm0.19$         \nl
 3C 437    & 1.48  & $11.4\pm0.8$  & $6.9\pm0.4$  & $4.2\pm0.2$   & $4.6\pm0.8$
 & $0.2\pm0.1$   & $46.18\pm0.10$         \nl
           &       & $10.5\pm0.8$  & $9.4\pm0.6$  & $6.7\pm0.4$   & $6.0\pm1.0$
 & $0.2\pm0.1$   & $46.49\pm0.10$         \nl

\enddata
\tablenotetext{a}{Lobe radius, 10 $h^{-1}$ kpc behind
hot spot, in $h^{-1}$ kpc. }
\tablenotetext{b}{Minimum energy magnetic field, 10 $h^{-1}$
kpc behind hot spot, in $h^{2/7} 10^{-5}$G.}
\tablenotetext{c}{Minimum energy magnetic field, 25 $h^{-1}$
kpc behind hot spot, in $h^{2/7} 10^{-5}$G.}
\tablenotetext{d}{Lobe advance speed, in $10^{-2}c$.}
\tablenotetext{e}{Ambient gas density, in $10^{-3}
h^{1/2} \mbox{cm}^{-3}$.}
\tablenotetext{f}{Logarithm of the luminosity in directed kinetic
energy, in $h^{-2} \mbox{erg s}^{-1}$.}
\end{deluxetable}

\clearpage

\clearpage

\clearpage

\addtocounter{figure}{-1}
\begin{figure}
\plotone{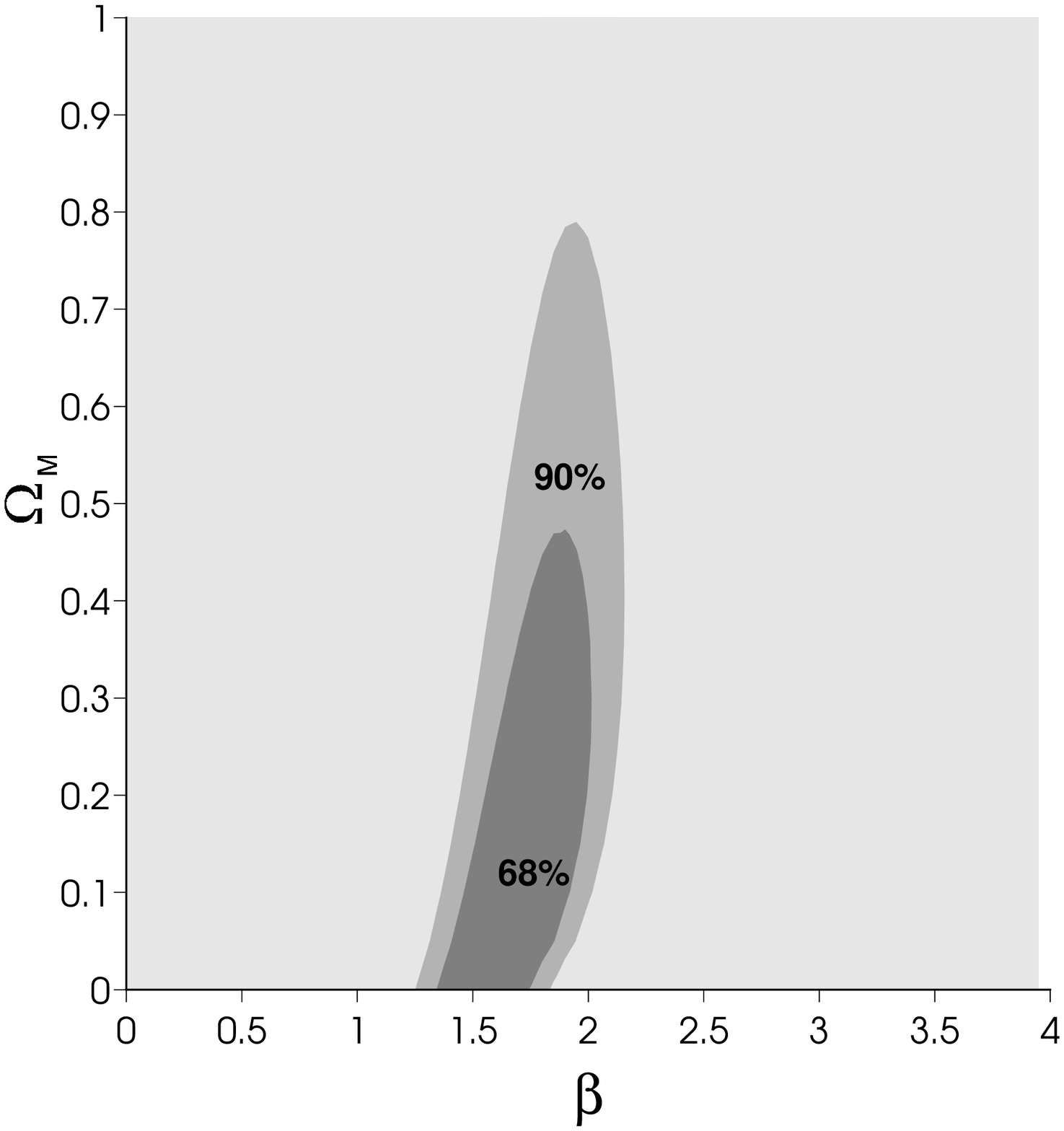}
\caption{...and (b) $\Omega_{\Lambda}=1-\Omega_m$ (spatially flat universe).} 
\end{figure}

\clearpage

\begin{figure}
\epsscale{0.75}
\plotone{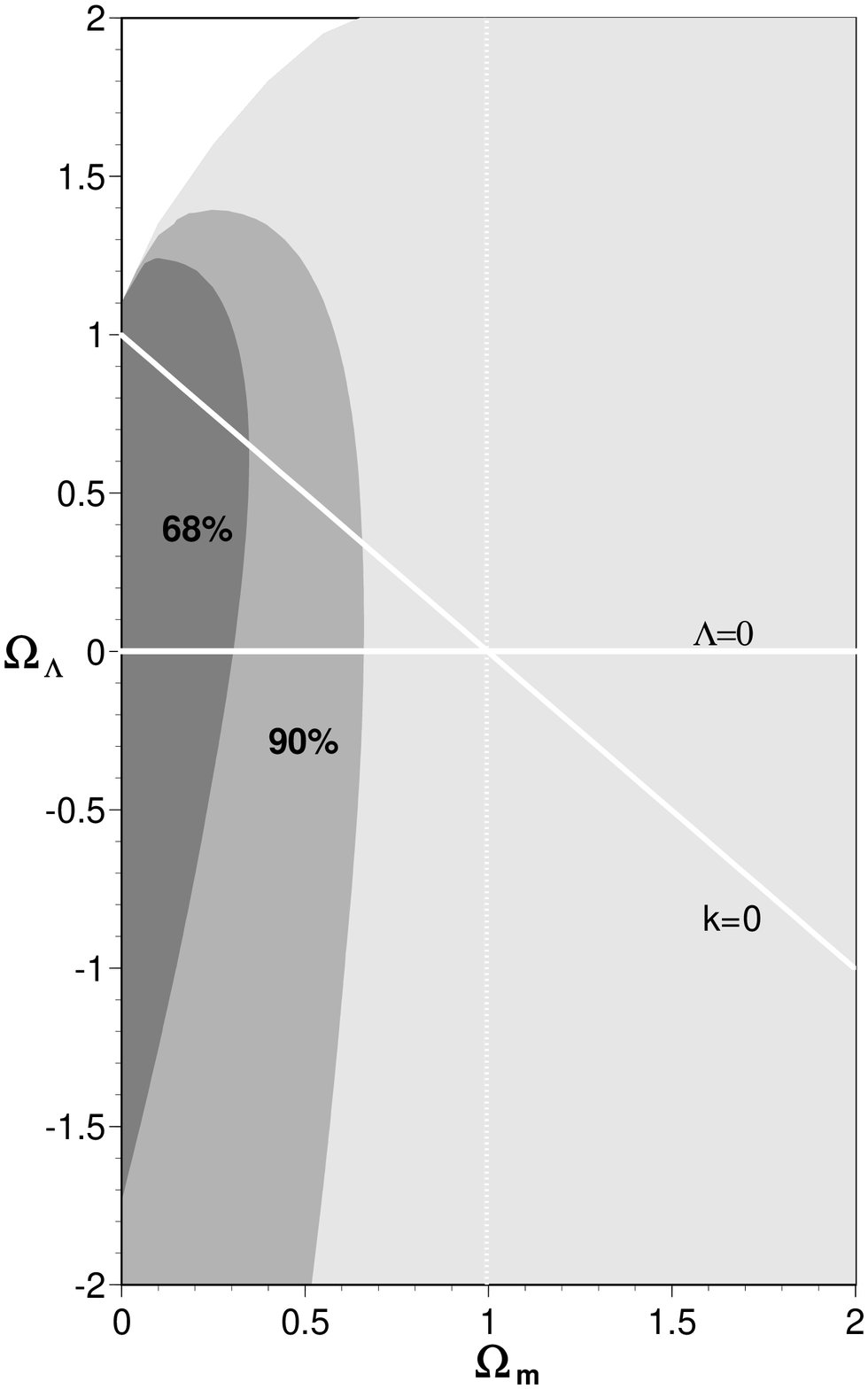}
\caption{The 68\% and 90\% confidence intervals for ranges of both
$\Omega_m$ and $\Omega_{\Lambda}$, independent of $\beta$. (Two-dimensional)
\label{fig:2d}}
\end{figure}

\clearpage

\begin{figure}
\epsscale{0.75}
\plotone{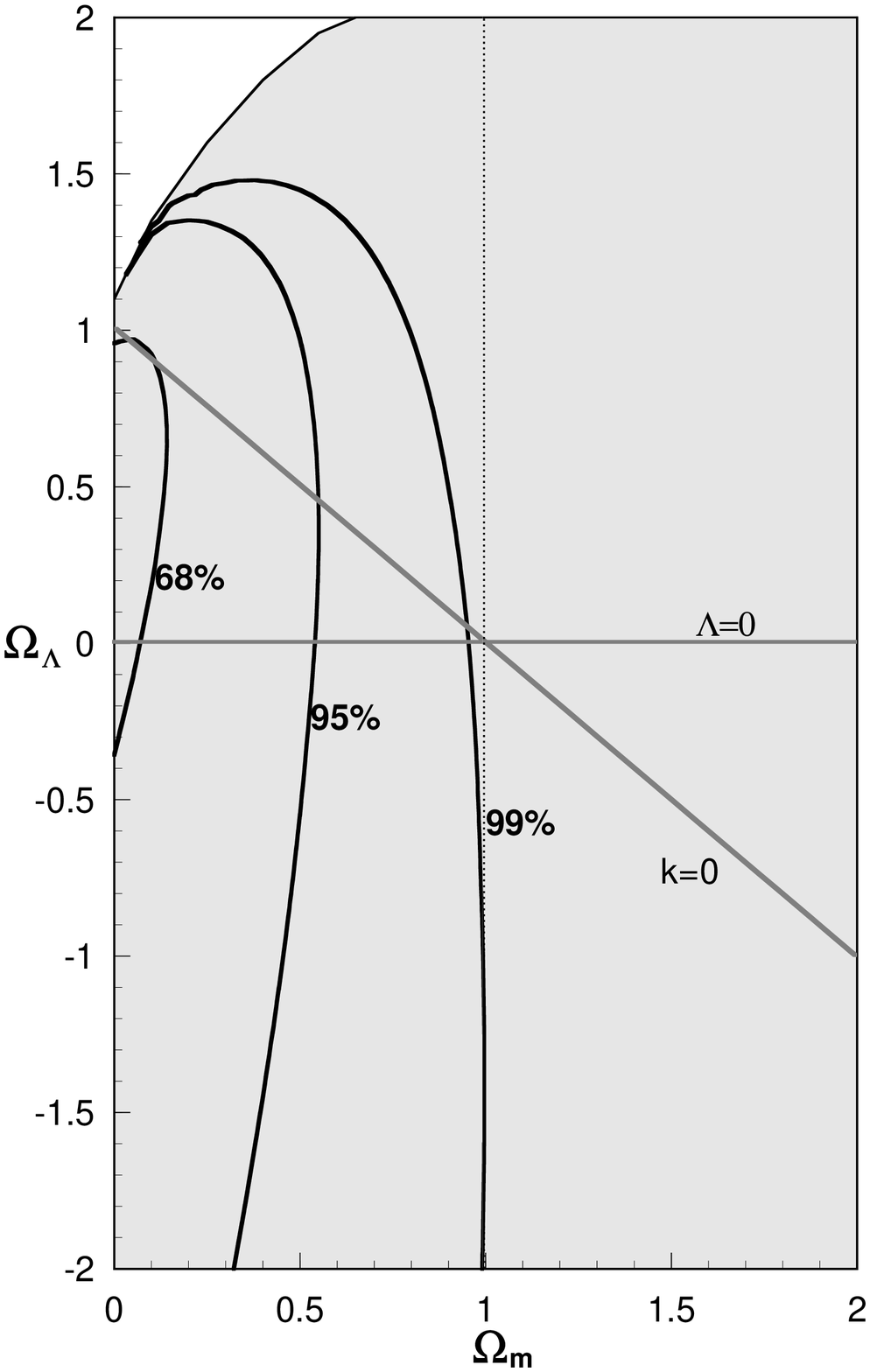}
\caption{The projections of the
68\%, 95\%, and 99\% confidence intervals
onto either axis ($\Omega_m$ or $\Omega_{\Lambda}$) 
indicates the probability associated with the
range in that one parameter, independent of
all other parameter choices. (One-dimensional) \label{fig:1d}}
\end{figure}

\end{document}